\documentclass[a4paper,fleqn,usenatbib]{mnras}
\usepackage{newtxtext,newtxmath}
\usepackage[T1]{fontenc}
\usepackage{ae,aecompl}
\usepackage{enumerate}
\usepackage{soul} 
\usepackage{lineno}
\usepackage{lipsum}
\usepackage{widetext}
\usepackage{tikz}
\def\checkmark{\tikz\fill[scale=0.4](0,.35) -- (.25,0) -- (1,.7) -- (.25,.15) -- cycle;}

\usepackage{comment}
\usepackage{graphicx}	
\usepackage{amsmath}	
\usepackage{amssymb}	
\usepackage{ulem}
\usepackage{color}

\def\apj{ApJ}                 
\def\apjl{ApJL}               
\def\apjs{ApJS}               
 
\def\mnras{MNRAS}             
\def\aap{A\&A}                
\def\nat{Nature}              
\def\araa{ARA\&A}             
\def\pasj{PASJ}               
\def\prl{Phys. Rev. Lett.}    

\def\e{\,\,\,\,}

\title{Radio Forensics Could Unmask Nearby Off-axis Gamma-ray Bursts}

\author[Bartos et al.]{I. Bartos\thanks{E-mail: imrebartos@ufl.edu},$^{1}$
K.H. Lee,$^{1}$
A. Corsi,$^{2}$
Z. M\'arka$^{3,4}$
S. M\'arka$^{3,4}$
\\
$^{1}$Department of Physics, University of Florida, Gainesville, FL 32611, USA\\
$^{2}$Department of Physics and Astronomy, Texas Tech University, Lubbock, TX 79409, USA\\
$^{3}$Columbia Astrophysics Laboratory, Columbia University, New York, NY 10027, USA \\ 
$^{4}$Department of Physics, Columbia University, New York, NY 10027, USA}

\begin{document}
\label{firstpage}
\pagerange{\pageref{firstpage}--\pageref{lastpage}}
\maketitle

\begin{abstract}
The multi-messenger observation of gamma-ray burst (GRB)\,170817A from the nearby binary neutron-star merger GW170817 demonstrated that low-energy $\gamma$-ray emission can be observed at relatively large angles from GRB jet axes. If such structured emission is typical, then the currently known sample of short GRBs with no distance measurements may contain multiple nearby off-axis events whose delayed afterglows could have gone undetected. These nearby neutron star mergers may produce telltale radio flares peaking years after the prompt GRB emission that could still be observable. Here, we show that several short GRBs observed by the Burst Alert Telescope (BAT) on the Neil Gehrels \textit{Swift} satellite, with no identified afterglow and no distance measurement, could potentially be associated with radio flares detectable by sensitive cm-wavelength radio facilities such as the Karl G. Jansky Very Large Array. We also examine optical follow-up observations that have been carried out for these events, and find that a nearby GW170817-like kilonova is ruled out for only a third of them.
\end{abstract}

\begin{keywords}
gamma-ray burst: general -- radio continuum: transients -- gravitational waves
\end{keywords}

\section{Introduction}
\label{sec:intro}
The remarkable multi-messenger discovery of the binary neutron star (BNS) merger GW170817 by the LIGO/Virgo gravitational-wave (GW) detectors and partner electromagnetic (EM) observatories, has marked the start of a new era in astrophysics. In addition to crowning multi-messenger astrophysics, confirming the predictions of Einstein's general theory of relativity, and clarifying the origin of the heaviest elements of the periodic table, GW170817 has given us new insight into the angular structure of relativistic jets in gamma-ray bursts (GRBs; e.g., \citealt{PhysRevLett.119.161101,2017ApJ...848L..12A,2017Sci...358.1556C,2017Sci...358.1583K,2017Sci...358.1579H,2017PASJ...69..102T,2017ApJ...848L..17C,2017ApJ...851L..21V,2017ApJ...848L..21A,2017ApJ...848L..20M,2017Natur.551...71T,2017Sci...358.1559K,2018A&A...613L...1D,margutti2018binary,2018Natur.554..207M,2017ApJ...850L..35A,2018ApJ...858L..15D,2018PhRvL.120x1103L,2018ApJ...861L..10C,2017arXiv171005931M} and references therein). 

At about $40$\,Mpc from Earth, GW170817 was associated with the closest and least energetic short GRB we know of (GRB\,170817A; \citealt{2017Sci...358.1556C,2017ApJ...848L..14G}). Its delayed afterglow, first detected in X-rays $ 9$ days after the merger \citep{2017Natur.551...71T,2017ApJ...848L..25H}, and in radio  $15$ days after the merger \citep{2017Sci...358.1579H}, further indicated its unusual nature. These observations suggested that the GRB was off-axis---its jet pointed away from Earth. 
Continued X-ray/optical/radio follow-up enabled a detailed reconstruction of the properties of the relativistic outflow of GRB\,170817A \citep{2018arXiv180502870A,2018ApJ...853L...4R,2018arXiv180609693M,2018PhRvL.120x1103L}. Overall, the emerging picture is that GRB\,170817 produced {\it structured} ejecta with a central, energetic jet core of half opening angle $\theta_{\rm j}\sim10^\circ$ \citep{2018ApJ...859L..23P}, plus fainter emission that extended to larger angular scales (tens of degrees) from the jet axis. The cause of this structured emission is not yet clear. It could originate from radiation from a very energetic but inefficient (in radiating gamma-rays) cocoon surrounding the ejecta \citep{2018Natur.554..207M,2018arXiv180609693M,2018ApJ...867...18N}, or a steep angular structure around the core, where the energy content at large angles is strongly diminished but the efficiency remains high \citep{2017ApJ...848L...6L,2018arXiv180806617V,2017ApJ...848L..20M}. Structured emission has been previously suggested based on the observed spectral breaks in afterglow emission \cite{2002MNRAS.332..945R} and the observed properties of prompt GRB emission \citep{2004ApJ...614L..13E}. At the same time the non-detection of a short GRB brighter than any of the archival events limits the range of opening angles \cite{2019MNRAS.483..840B}. It is worth noting that similar gamma-ray emission at large angles is likely strongly suppressed in long GRBs, as indicated from the comparison of their prompt and afterglow properties and by considering the shape of their luminosity function \citep{2019MNRAS.482.5430B}. 

Prior to GRB\,170817A, observations of so-called jet breaks in the X-ray and optical afterglow light curves of several GRBs indicated that these explosions are highly beamed  \citep{2006RPPh...69.2259M}. Typical reconstructed half opening angles for short GRBs are around $\theta_{\rm j}\sim10^\circ$ \citep{2014ARA&A..52...43B,2015ApJ...815..102F}.  Nonetheless, evidence for more structured jets had emerged in some cases (e.g., \citealt{2003ApJ...594..379P}). Some afterglow observations suggested a two component jet with an inner, narrow ultra-relativistic core, and an outer, wider and mildly relativistic component \citep{2000ApJ...538L.129F,2003Natur.426..154B,2003ApJ...595L..33S}. Structured emission was also motivated theoretically \citep{lipunov2001gamma,zhang2002gamma,pescalli,2005MNRAS.360..305S}.

The structure of short GRB jets has important consequences for the electromagnetic observability and multi-messenger detection rate of BNS mergers. In particular, sufficiently nearby GRBs may be detected in $\gamma$-rays even if their luminous jet core is not beamed towards Earth (off-axis jet), due to the weaker emission from the jet wings \citep{2017MNRAS.471.1652L,2018PhRvL.120x1103L,2018Natur.554..207M,2018PTEP.2018d3E02I,2017ApJ...850L..24G}. This makes the local BNS merger population over-represented within the detected sample of GRBs. Taking the structured emission of GRB\,170817A as standard, \cite{gupte} found that $\sim 10\%$ of observed short GRBs could  be associated with BNS mergers within 200\,Mpc, although this fraction could be less if GRB\,170817A is atypical \citep{2018arXiv180804831B}. A large fraction of local short GRBs has also been inferred from their directional correlation with nearby galaxies \citep{2005Natur.438..991T}. 

Detecting the afterglow of a potential population of nearby, off-axis short GRBs may not be as straightforward as for the case of on-axis events. Indeed, off-axis afterglow emission peaks later than typically expected for on-axis GRBs, and larger viewing angles result in weaker afterglows \citep{2002MNRAS.332..945R,2003ApJ...591.1075K,2003ApJ...591.1086G,2010ApJ...722..235V,2019arXiv190105792R}. Kilonova emission could be more promising as it is isotropic and may be detected for about a week after the merger from nearby events  \citep[or even more distant ones if accurate localization enables deep follow-up;][]{2013Natur.500..547T,2018NatCo...9.4089T}. In fact, the lack of such a kilonova signature in deep follow-up observations of nearby GRB candidates has been used to constrain the nature of GRB progenitors and/or to set a lower limit on the actual GRB distance under the hypothesis of a binary merger origin  \citep[see e.g.,][and Section \ref{sec:sample}]{2008MNRAS.384..541L}. 

Before GW170817, short GRBs detected by \textit{Swift} lacking arcsec localizations have not been monitored systematically for kilonova emission, leaving many of them unconstrained. Thus, the question of whether one could still identify these events as nearby ($\lesssim 200$,Mpc) long after the GRB explosion, has gained renewed interest. Here, we explore whether
radio remnants of BNS mergers may be the Rosetta Stone that establishes the nearby origin of off-axis GRBs, even long after their observation in $\gamma$-rays. During the merger, NSs eject some of their matter through tidal disruption and winds. The interaction of this mildly-relativistic matter with the circum-merger medium may give rise to radio emission that can last for years \citep{2011Natur.478...82N}. This very-late-time radio emission could be identified out to possibly 200-300\,Mpc, depending on the density $n$ of the circum-merger  medium \citep{2016ApJ...831..190H,2019arXiv190100868K}. 
Searches for radio remnants of short GRBs carried out so far have targeted only well localized events, all located at $>500$\,Mpc \citep{2016ApJ...819L..22H,2016ApJ...831..141F}. These searches have resulted in non-detections only, and were focused on the search for long-lived magnetar remnants from the binary merger. In this paper we show that, if GRB ejecta are structured similarly to GW170817, then radio observatories such as the Karl G. Jansky Very large Array (VLA) may help us identify nearby events in the sample of previously detected short GRBs with no observed afterglows and unknown distances. 

Our discussion is organized as follows. In Section \ref{sec:rate} we discuss the expected detection rate of nearby short GRBs and show that there should be some nearby GRBs among those detected so far. In Section \ref{sec:whynodetection} we review why these nearby GRBs could have gone unidentified as local events. In Section \ref{sec:sample} we introduce \textit{Swift}/BAT GRBs of interest that have no identified afterglows. We also discuss constraints on their distances from previous optical observations that rule out a nearby kilonova (for some of the bursts). In Section \ref{sec:promisingtargets} we calculate the expected radio remnant flux for GRBs in our sample, discuss their detectability with the VLA, and their expected time evolution. In Section \ref{sec:galaxies} we comment on directionally overlapping nearby galaxies for the same GRBs. We conclude in Section \ref{sec:conclusion}. 

\section{Expected detection rate of nearby short GRB\lowercase{s}}
\label{sec:rate}
For structured jets like GRB\,170817A, a GRB can be observed even off-axis---at viewing angles $\theta_{\rm v} > \theta_{\rm j}$---as long as the GRB is sufficiently nearby, where the close distance compensates for the weak emission \citep{2017MNRAS.471.1652L,2018PhRvL.120x1103L,2018Natur.554..207M,2018PTEP.2018d3E02I,2017ApJ...850L..24G}. Observability out to greater $\theta_{\rm v}$ means a higher rate of detection. Therefore, there should be more observed nearby GRBs than expected from the rate of distant GRBs, and nearby GRBs should typically be observed at greater $\theta_{\rm v}$ than distant GRBs. The reconstructed large viewing angle of $14^\circ\lesssim\theta_{\rm v}\lesssim 40^\circ$ from the nearby GRB\,170817A is in line with this expectation \citep{2018arXiv180511579T,2018ApJ...860L...2F}.

If GRB\,170817A is typical, then similar, nearby short GRBs with large $\theta_{\rm v}$ should have been detected in the past. Considering a detection threshold fluence of $2.5 \times 10^{-8}$\,erg\,cm$^{-2}$ and GRB\,170817A's fluence of $\approx 3\times 10^{-7}$\,erg\,cm$^{-2}$, both for {\it Fermi}-GBM in the $10-1000$\,keV energy range \citep{2017ApJ...848L..14G}, a GRB\,170817A-like event could be observed out to a distance of $\sim130$\,Mpc. This distance grows further for smaller $\theta_{\rm v}$. 

To estimate the expected number of detected nearby short GRBs, \cite{gupte} considered the structured jet model of \cite{margutti2018binary}. They took the local BNS merger rate of $1540^{+3200}_{-1220}$\,Gpc$^{-3}$yr$^{-1}$ inferred from LIGO/Virgo observations \citep{PhysRevLett.119.161101}, and considered that \textit{Fermi} (\textit{Swift}) monitors 70\% (15\%) of the sky at any given time. They found that about 10\% of detected short GRBs from BNS mergers should have occurred within 200\,Mpc, the average distance out to which GW observatories will be able to detect BNS mergers when reaching full sensitivity \citep{2018LRR....21....3A}. Considering that the total short GRB detection rate of \textit{Fermi}/GBM and \textit{Swift}/BAT is 40\,yr$^{-1}$ \citep{2016ApJS..223...28N} and 8\,yr$^{-1}$, respectively, up to tens of previously observed GRBs could be of local origin. \cite{2019MNRAS.483..840B} used the observed short-GRB luminosity function to calculate the expected fraction of nearby, off-axis short GRBs. They find a somewhat smaller fraction, $1\%-10\%$ for these events, and conclude that GRB\,170817 is probably unusual.

\section{Could nearby short GRB\lowercase{s} go unrecognized?}
\label{sec:whynodetection}

Despite the non-negligible expected rate of nearby ($\lesssim 200$\,Mpc) short GRBs, none had been identified prior to GW\,170817A, and GRB\,170817A itself would probably not have been recognized as a nearby event without the GW detection motivating an aggressive broad-band follow-up campaign in its relatively large sky localization area \citep{2017ApJ...848L..12A,2017Sci...358.1556C}. Indeed, $\gamma$-rays alone mostly cannot provide accurate enough localization to allow for host galaxy identification and distance measurement. Detecting GRB afterglows, which could provide arcsec localizations, is only straightforward for on-axis events, for which the X-ray afterglow is the brightest shortly after the GRB, following which it gradually fades \citep{2009MNRAS.397.1177E,2006ApJ...642..389N}. The X-ray afterglow of an off-axis GRB, on the other hand, is weaker at peak compared to an on-axis event, and the peak time is delayed possibly by days after the GRB \citep{2010ApJ...722..235V,2011ApJ...733L..37V}. So far, within the sample of short GRBs, a delayed X-ray afterglow has been detected for only two bursts: GRB\,150101B, whose X-ray afterglow was fortuitously identified thanks to its vicinity to a low-luminosity AGN initially mistaken for the candidate X-ray counterpart \citep{2018arXiv180610624T}; and GRB\,170817A, for which the GW localization led to the identification of the optical kilonova to arcsec position, thus allowing X-ray/radio follow-up.

The optical kilonovae that are produced in the aftermath of BNS mergers by dynamical and wind ejecta last for about a week, with optical/near-infrared emission of $10^{40}-10^{42}$\,erg\,s$^{-1}$ \citep{2017LRR....20....3M,2017ApJ...848L..17C}. Although kilonovae have the advantage of being potentially observable regardless of viewing geometry, their optical/NIR luminosity makes it challenging for most current large optical surveys to discover them \citep{2017ApJ...851L..48Y,2017ApJ...848L..17C,2018ApJ...852L...3S}. This will likely change in the future with the completion of large-scale survey instruments such as the Large Synoptic Survey Telescope \citep{2018arXiv181103098C}.
Kilonova emission can be more easily identified if a GRB is detected in association with it, thus enabling more accurate localization \citep{2013Natur.500..547T}. While the arcmin-level localizations available for short GRBs without an observed X-ray afterglow can still challenge the identification of a kilonova at large distances \citep{2015ApJ...806...52S}, a kilonova from nearby GRBs with arcmin localizations could be detected using systematic, targeted follow-up with meter-class telescopes (see Appendix and Table \ref{table:GRBs}). 

As discussed in Section \ref{sec:intro}, slow ejecta from BNS mergers should also produce radio flares by interacting with the surrounding medium. These flares are the focus of this paper. They can last for years after the merger \citep{2011Natur.478...82N,2016ApJ...831..190H}, and could be detectable from nearby ($\lesssim 200$\,Mpc) events, depending on the ejecta velocity distribution and the density of the circum-merger medium. However, typically, only short GRBs localized to arcsec positions are extensively followed up in the radio, and all of these well localized short GRBs (other than GRB\,170817A) reside beyond $\sim 1$\,Gpc, which makes it unlikely for a radio flare from slow BNS ejecta to be detectable. Hereafter, we explore the possibility of detecting very-late-time radio flares from BNS ejecta associated with nearby short GRBs lacking an X-ray afterglow identification (possibly because off-axis), and localized to arcmin positions only thanks to their $\gamma$-ray emission.  

\section{Nearby GRB candidates}
\label{sec:sample}
To identify promising candidates for nearby GRBs, we considered short GRBs in the \textit{Swift} catalog with no afterglow detection and thus no known distance measurement. We only considered short GRBs detected by \textit{Swift}/BAT as they have 90\% error radii smaller than half the FWHM of the VLA primary beam at 6\,GHz ($\approx 7.5$\,arcmin), making them suitable to be covered with a single VLA pointing. We excluded from the sample GRBs with Declination below $-40^\circ$ as those would not be observable with the VLA, and also GRBs with declinations in the range Dec=$[-5^\circ,15^\circ]$ as observations of them can be significantly degraded due to satellite transmission. We list the remaining, suitable GRBs in Table \ref{table:GRBs}. 

\subsection{Constraints from previous observations}

Among other properties, in Table \ref{table:GRBs} we list the $\gamma$-ray fluence of GRBs in our sample as measured by \textit{Swift}/BAT. The measured fluences range from $f_{\gamma} = (1.2\times10^{-8}-2.6\times10^{-7})$\,erg\,cm$^{-2}$ in the $15-150$\,keV band, typical for short GRBs detected by \textit{Swift}/BAT \citep{2016ApJ...829....7L}. Considering a fiducial distance of 200\,Mpc, these fluences can be converted to isotropic-equivalent $\gamma$-ray energies as (see Eq. 6 of \citealt{2015ApJ...815..102F}):
\begin{equation}
E_{\rm \gamma,iso} = k_{\rm bol} \times \frac{4\pi d_{\rm L}^2}{1+z}f_{\gamma},
\end{equation}
where $k_{\rm bol}=5$ is the bolometric correction factor to convert the fluence to an energy range of $\approx 1-10^4$\,keV, $d_{\rm L}=200$\,Mpc is the luminosity distance, and $z$ is the redshift. Using this conversion we obtain a range of $E_{\rm \gamma,iso}=3\times10^{47}-6\times10^{48}$\,erg. For comparison, on axis short GRBs are measured to have $E_{\rm \gamma,iso}=3\times10^{49}-3\times10^{52}$\,erg (e.g., Fig 7 of \citealt{2015ApJ...815..102F}), while GRB\,170817A was measured to have $E_{\rm \gamma,iso}\approx3\times10^{46}$\,erg. This implies that it is unlikely that GRBs in our sample on-axis bursts located at $d_L\lesssim 200$\,Mpc (see also \citealt{2019MNRAS.483..840B}). However, their fluences are consistent with both nearby but off-axis GRBs, and distant but on-axis events. 

We further examined whether optical follow-up observations of these GRBs---which were primarily aimed at identifying their afterglow emission---can set constraints on their distances assuming they are all associated with GW170817-like isotropic kilonovae. Specifically, we considered the observed kilonova light curve of GW170817, and checked for each GRB whether the available optical follow-up observations could have uncovered such a kilonova at a distance of $d_L\lesssim 200$\,Mpc. We found that about a third of the GRBs in our sample are constrained by these observations to be either farther than 200\,Mpc, or not associated with a GW170817-like kilonova. We mark these GRBs in the last column of Table \ref{table:GRBs} as they are not of high interest for a search of nearby radio flares as described in Section \ref{sec:promisingtargets} (see the Appendix for further details).

Finally, Table \ref{table:GRBs} also lists available exclusion distance lower limits from astrophysically triggered searches \citep{2008CQGra..25k4051A} using GW data. Of the GRBs in our sample, only GRB\,151228A has a measured exclusion distance lower limit of 122\,Mpc for a BNS merger, which sets a lower distance limit for this event \citep{2017ApJ...841...89A}.

\subsection{Spatial correlation with local galaxies}
\label{sec:galaxies}
As an independent check for nearby GRBs in our sample, we searched for spatially coincident galaxies with luminosity distances below $200$\,Mpc using a publicly available galaxy catalog \citep{2018MNRAS.479.2374D}. The catalog has a stated completeness of 40\% in blue luminosity within 200\,Mpc, therefore a lack of overlap does not rule out the local origin of a GRB. 

We searched for galaxies within 200\,Mpc in the catalog whose angular distance from the direction of GRBs in our sample is less than the sum of the GRB's 90\% error radius and an offset corresponding to a projected physical offset of $30$\,kpc. The projected offset was calculated for every galaxy individually. This latter term takes into account that BNS systems can move away from their host galaxies due to natal kicks experienced by neutron stars at their formation. We use 30\,kpc as it includes $\sim90$\% of observed projected physical offsets for short GRBs \citep{2014ARA&A..52...43B}.
We found that GRB\,050906 has two and spatially coincident galaxies within 200\,Mpc, at 130\,Mpc and 150\,Mpc from Earth. While in the sky area contaminated by satellite transmission, we note that GRB\,070810B also has a directionally coincident galaxy at 170\,Mpc from Earth.

To determine the probability of a chance coincidence, we counted the number of galaxies within 200\,Mpc of Earth and within 2$^\circ$ of the direction of GRBs in our sample, but outside of the 90\% error radius plus 30\,kpc offset. We used this number for each GRB to estimate the expected number of chance coincidences.
We found that the probability of observing each galaxy as a chance coincidence for GRB\,050906 and GRB\,070810B is of $9\%$ and $5\%$, respectively. Therefore, these galaxy overlaps by themselves do not establish the local origin of GRB\,050906 and GRB\,070810B. Moreover, optical non-detections rule out a GW170817-like kilonova at $d_L\lesssim 200$\,Mpc for both these events (see Table \ref{table:GRBs}).

\section{Identifying nearby short GRB\lowercase{s} with radio forensics}
\label{sec:promisingtargets}

Here, we are interested in testing whether any of the GRBs in our sample could be identified as nearby by means of the very late-time radio flares expected to be associated with BNS slower ejecta. We make the assumptions that all GRBs in our sample produce slow ejecta with properties similar to the slow ejecta that powered the GW17017 kilonova. Under this assumption, we discuss the predictions for the expected luminosity of the associated radio flares (Section \ref{sec:expected}). Then, we present the practical implementation of a search for late-time radio flares with the VLA (Section \ref{sec:detectability}).

\begin{figure}
\centering
\includegraphics[width=0.49\textwidth]{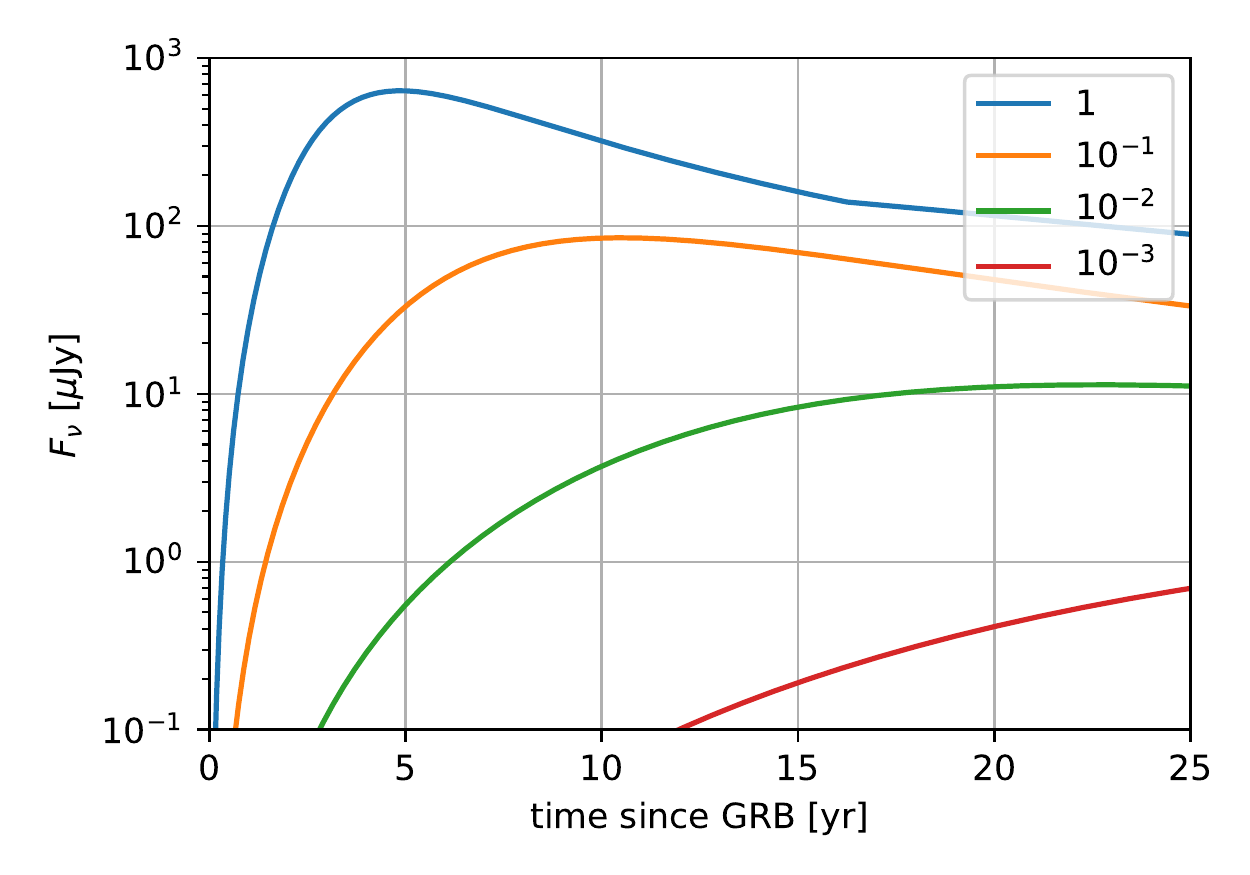}\\
\caption{Expected 6\,GHz radio flux from a BNS ejecta for a source at 200\,Mpc as a function of time after merger, for different circum-merger densities (see legend; densities are in units of cm$^{-3}$). The ejecta was assumed to have a fast component with $0.01$\,M$_{\odot}$ mass expanding with $0.3c$ speed, and a slow component with $0.04$\,M$_{\odot}$ mass expanding with $0.1c$ speed.}
\label{fig:radioflux}
\end{figure}

\subsection{Expected radio emission}
\label{sec:expected}
We calculate the radio flare light curve expected for different BNS circum-merger densities. We adopt an outflow compatible with the reconstructed properties of of the kilonova associated with GW170817. These properties indicated the presence of two components: a faster component with $0.01$\,M$_{\odot}$ mass and $0.3c$ velocity, and a slower component with $0.04$\,M$_{\odot}$ mass and $0.1c$ velocity \citep{2017ApJ...848L..17C}. We assumed that all BNS ejecta have these two components.

We calculate the expected radio fluxes following the prescriptions of \cite{2013MNRAS.430.2121P}. We consider an expanding spherical, non-relativistic ejecta that collects mass from the uniform circum-merger medium with particle density $n$ as it moves outwards with velocity $v\equiv\beta c$, where $c$ is the speed of light. The velocity of the ejecta decreases as it collects more mass. The relation between ejecta velocity and radius ($R$) can be calculated by assuming that kinetic energy is conserved (see Eq. 14 of \citealt{2013MNRAS.430.2121P}):
\begin{equation}
M(R)(\beta c)^2 \approx E(\geq\beta),
\end{equation}
where $M$ is the collected mass at radius $R$ from the merger, and $E(\geq\beta)$ is the kinetic energy of the part of the ejecta that was launched with velocities faster than $\beta c$. In our case, this means that initially only the component with $0.3c$ velocity affects the evolution of the ejecta front and hence the radio flux until it slows down to $0.1c$ where the slower component can catch up. With $R$ and $\beta$ known functions of time, we can calculate the emitted radio flux as (see Eqs. 4 and 6 and Table 2 in \citealt{2013MNRAS.430.2121P}) 
\begin{equation}
F(t)\approx 4\,\mbox{mJy}\,R_{17}^3 n_{-1}^{15/8} \beta^{19/4} d_{200}^{-2}\left(\frac{\nu}{6\,\mbox{GHz}}\right)^{-3/4},
\end{equation}
where $R_{17}\equiv (R/10^{17}\,\mbox{cm})$, $n_{-1}\equiv (n/0.1\,\mbox{cm}^{-3})$, and $d_{200}\equiv(d/200\,\mbox{Mpc})$. We assume that the observational frequency $\nu_{\rm obs}=6$\,GHz is greater than both the typical synchrotron frequency of electrons in the forward shock driven by the ejecta ($\nu_{\rm m}$), and the synchrotron self-absorption frequency ($\nu_{\rm a}$; see e.g.,  \citealt{2013MNRAS.430.2121P}). We further assume that electrons and the magnetic field carry $\epsilon_{\rm e}\approx\epsilon_{\rm m}\sim10\%$ of the total internal energy of the shocked gas, and that the power-law index of the distribution of the accelerated electrons' Lorentz factors is $p=2.5$. 

Our estimated radio light curves are shown in Fig. \ref{fig:radioflux} for different circum-merger densities. Our results are similar to those of \cite{2013MNRAS.430.2121P}, who simulated the mass ejection from BNS mergers and found that most mass will be ejected at $\sim0.3c$. Our flux is also dominated by the $0.3c$ component; the slower component only plays a role on longer time scales than the first two decades after the GRB we focus on here. Our results are also similar to those of \cite{2019arXiv190100868K} for their $0.3c$ component after scaling the results to the same distance and radio frequency. At the same time, our estimated flux evolution is significantly slower than that of \cite{2016ApJ...831..190H}. This is because \cite{2016ApJ...831..190H} assumed that much of the ejecta has a velocity higher than $0.3c$, which results in faster rise time and more rapid fading of the radio remnant.

\subsection{Identification with the VLA}
\label{sec:detectability}
Based on the predictions described in Section \ref{sec:expected}, we calculate what would be the expected  6\,GHz radio flux for observations of the short GRBs in our sample (see Section \ref{sec:sample}), and estimate the maximum distance to which this radio emission could be detected with VLA. 

We assumed an observing campaign where each GRB is observed over two epochs, once during the VLA semester 2019A (May 31, 2019; epoch I), and a second time during semester 2020A (May 31, 2020; epoch II). We chose these VLA observing semesters because they are the closest to the current date where the VLA is not in its most compact D configuration, which would pose challenges for disentangling any radio flare from possible contamination from the host galaxy \citep[e.g.,][]{2016ApJ...829L..28P}.  In fact, the VLA $\approx 1$\,arcsec resolution in B configuration and at 6\,GHz is ideal as at 200\,Mpc is corresponds to scale of $\approx 0.89$\,kpc. Thus, such resolution would be enough to disentangle radio emission from a radio flare located at 200\,Mpc, for offsets from its hosts $\gtrsim 3$\,kpc  (the last is comparable to that measured for GW170817, and closer than $90\%$ of offsets measured for cosmological short GRBs; \citealt{2017ApJ...848L..28L}). We assume that each observing epoch is 1\,hr long, which is enough for the VLA to reach an r.m.s. sensitivity of $\approx 3$\,$\mu$Jy at 6\,GHz in its B configuration (for a nominal bandwidth of 4\,GHz and accounting for RFI effects and calibration overhead).

\begin{figure}
\centering
\includegraphics[width=0.47\textwidth]{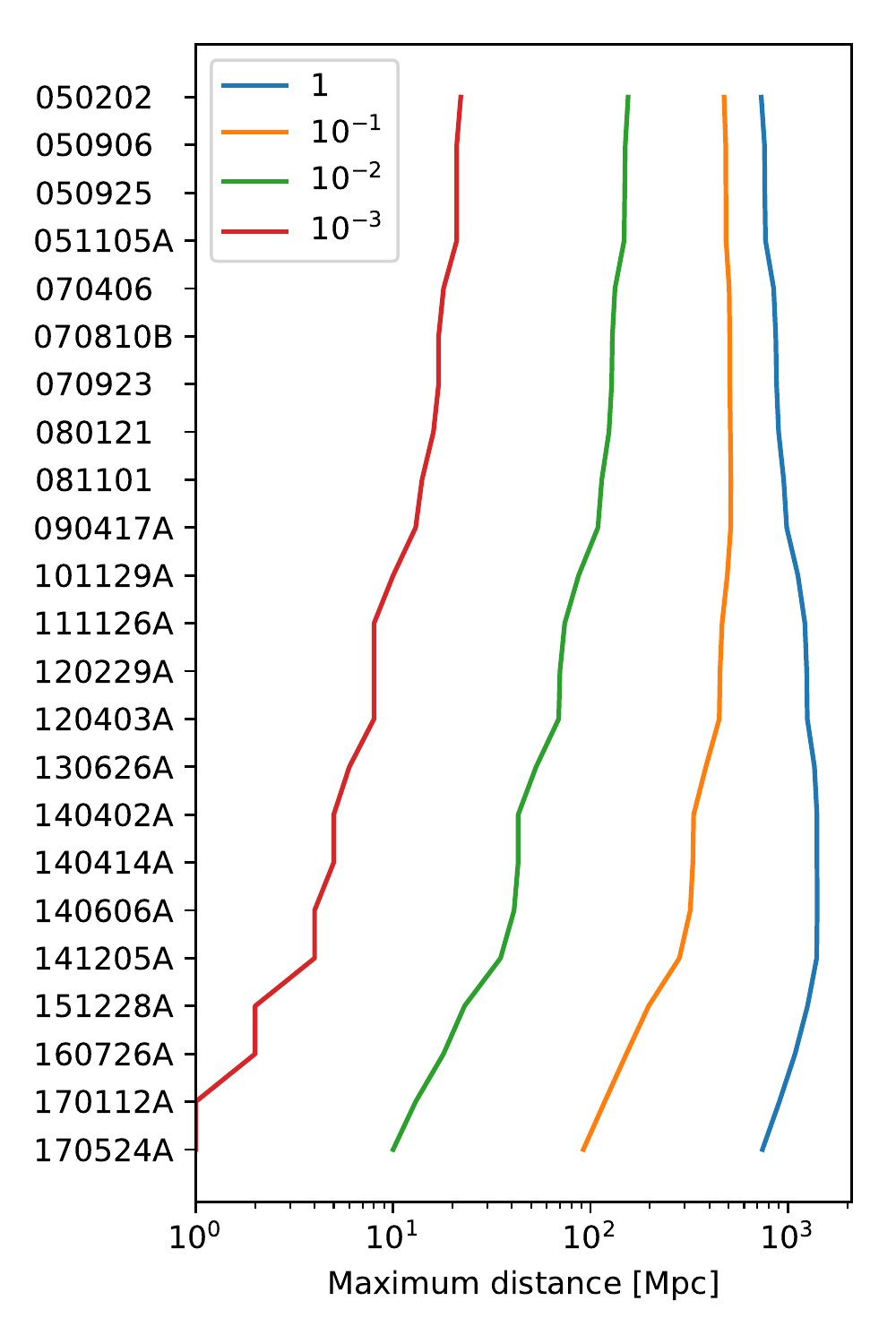}\\
\caption{Maximum distance out to which a VLA observation in Epoch I (see Section \ref{sec:detectability}) could be used to detect the radio remnant of \textit{Swift} short GRBs in our sample, for different circum-merger medium densities. \textit{Swift} GRBs used in this analysis are listed along the vertical axis. Circum-merger densities are indicated in the legend, in units of cm$^{-3}$.}
\label{fig:maxdistanceref}
\end{figure}

Fig. \ref{fig:maxdistanceref} shows the maximum distances to which radio emission is detectable during Epoch I, assuming that a pre-GRB radio image is available. For this case, we consider fluxes above $9\,\mu$Jy to be detectable. For high circum-merger densities, the radio remnant is detected for all GRBs out to large distances. For 1\,cm$^{-3}$ density this is roughly 1\,Gpc for all GRBs. However, for circum-merger densities $\lesssim10^{-3}$, which may represent as much as half of binary mergers \citep{2015ApJ...815..102F}, radio emission is not detectable for our model.

In cases for which no pre-GRB radio images are available, it is also of interest to evaluate whether any radio flux variability can possibly be detected over the two epochs, so it can be used as a discriminant against unrelated radio sources in the field. Specifically, we deem radio variability significant when $|F_{\rm I}-F_{\rm II}|/\sigma_{\rm (F_{\rm I}-F_{\rm II})}\geq 3$, where $\sigma_{\rm (F_{\rm I}-F_{\rm II})} $ is calculated accounting for both the statistical r.m.s. image error, and a $5\%$ absolute flux calibration error on each of the flux measurements, i.e.: 
\begin{equation}
\sigma_{\rm (F_{\rm I}-F_{\rm II})} \equiv \sqrt{(0.05F_{\rm I})^2 + (0.05F_{\rm II})^2 + 2\times(3\,\mu\mbox{Jy})^2}
\end{equation}

\begin{table*}
	\centering
	\caption{\label{table:GRBs} List of short GRBs detected by {\it Swift}/BAT with no identified afterglow and observable with the VLA (see text for discussion). The columns are: GRB name; Right Ascension; Declination; 90\% error radius of the localization; $\gamma$-ray fluence ($15-150$\,keV); delay between the GRB trigger and beginning of \textit{Swift}/XRT follow-up observation ("--'' indicates XRT no observations have been carried out; see more details and references in the Appendix);	published 90\% confidence lower limits on the exclusion distance for BNS mergers by targeted GW observations \citep{2010ApJ...715.1453A,2012ApJ...760...12A,2017ApJ...841...89A}; in the last column we marked GRBs for which optical follow-up observations rule out a GW170817-like kilonova within 200\,Mpc. } 
	\begin{tabular}{l|c|c|c|c|c|c|c|c|}
		\hline
\e GRB    & R.A.    & Dec.    & err.rad. & $f_{\gamma}$               & $t_{\rm XRT}$ & $D_{\rm GW}$ & GW170871-like kilonova at  \\ 
\e name   & [deg]   & [deg]   & [arcmin]   & [erg\,cm$^{-2}$] &             & [Mpc]        & $d_L\lesssim 200$\,Mpc excluded? \\\hline        \hline
170524A   & 319.488 &  48.603 & 2          & $3.7\times10^{-8}$ & 61\,s       &  & \checkmark \\ \hline 
170112A   & 15.232  &  17.233 & 2.5        & $1.3\times10^{-8}$ & 62\,s       &  &  \\ \hline 
160726A   & 98.809  &  -6.617 & 1.3        & $2.6\times10^{-7}$ & --          &  &  \\ \hline
151228A   & 214.017 & -17.665 & 1.8        & $8.4\times10^{-8}$ & --          &  122 &  \\ \hline 
141205A   & 92.859  &  37.876 & 2          & $1.2\times10^{-7}$ & 6.7\,h      &  &  \\ \hline
140606A   & 201.799 &  37.599 & 2.4        & $5.1\times10^{-8}$ & 1.3\,h      &  &  \\ \hline
140414A   & 195.31  &  56.902 & 4          & $1.2\times10^{-7}$ & 11.2\,h     &  & \checkmark \\ \hline
130626A   & 273.128 &  -9.525 & 1.8        & $5.2\times10^{-8}$ & 111\,s      &  &  \\ \hline
120403A   & 42.458  &  40.489 & 2.3        & $1.0\times10^{-7}$ & --          &  &  \\ \hline
120229A   & 20.033  & -35.796 & 1.9        & $4.1\times10^{-8}$ & --          &  &  \checkmark \\ \hline
111126A   & 276.057 &  51.461 & 3          & $7\times10^{-8}$ & --          &  &  \\ \hline
101129A   & 155.921 & -17.645 & 3          & $9\times10^{-8}$ & 11\,h       &  &  \\ \hline
090417A   & 34.993  &  -7.141 & 2.8        & $1.9\times10^{-8}$ & --          &  &  \\ \hline
080121\e  & 137.235 &  41.841 & 3          & $3\times10^{-8}$ & 2.3\,d      &  &  \\ \hline
070923\e  & 184.623 & -38.294 & 2.1        & $5.0\times10^{-8}$ & --          &  5.1 &  \\ \hline 
070406\e  & 198.956 &  16.53  & 3.3        & $4.5\times10^{-8}$ & 20.6\,h     &  &  \checkmark \\ \hline
051105A   & 265.279 &  34.916 & 2.6        & $2.0\times10^{-8}$ & 67\,s       &  & \checkmark \\ \hline
050925\e  & 303.49  &  34.329 & 1.4        & $7.5\times10^{-8}$ & 91\,s       &  &  \\ \hline
050906\e  & 52.802  & -14.621 & 3.3        & $5.9\times10^{-8}$ & 79\,s       &  & \checkmark \\ \hline
050202\e  & 290.584 & -38.73  & 2.3        & $3.0\times10^{-8}$ & --          &  &  \\ \hline
	\end{tabular}
\end{table*}

\begin{figure}
\centering
\includegraphics[width=0.47\textwidth]{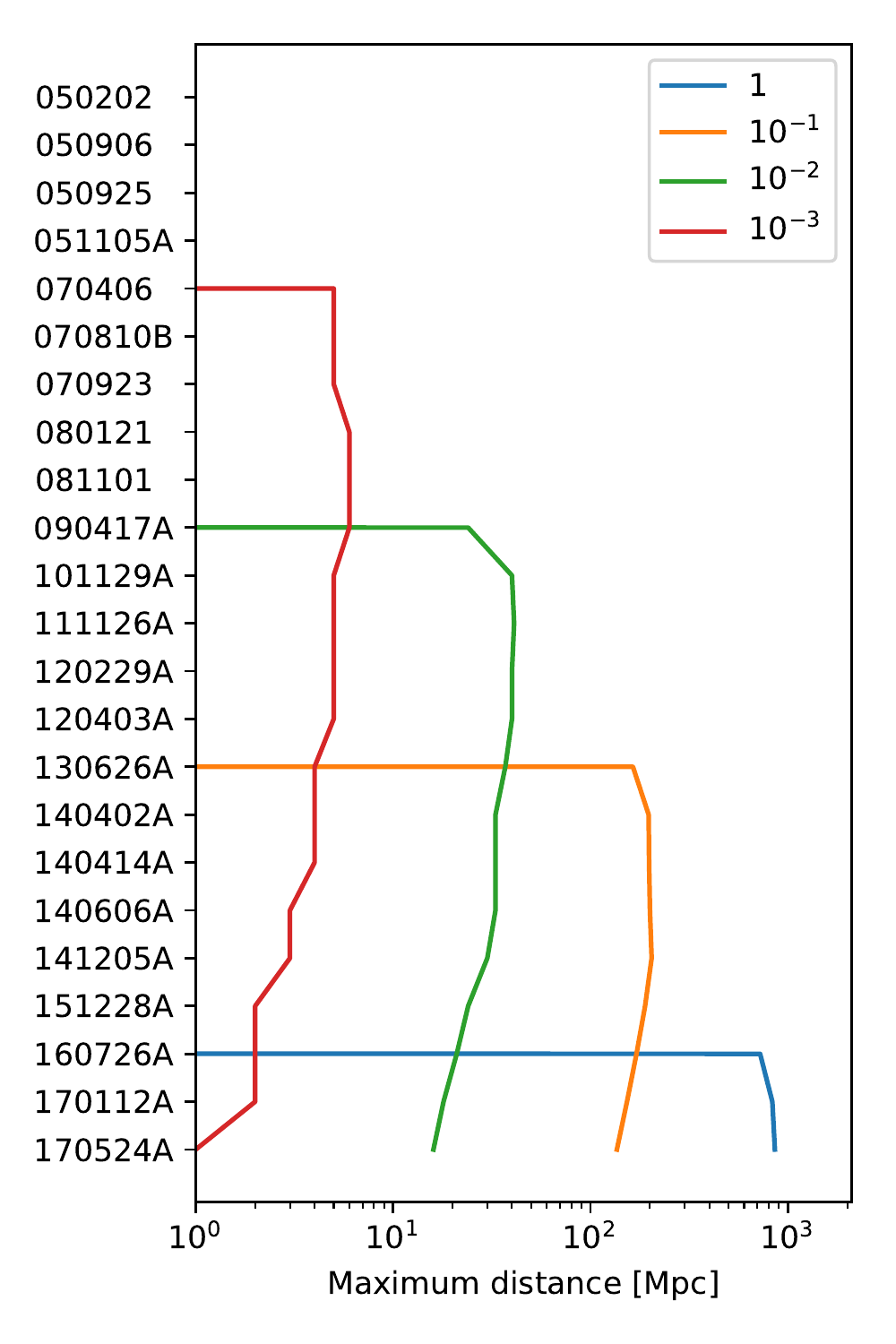}\\
\caption{Maximum distance out to which VLA observations in Epochs I and II (see Section \ref{sec:detectability}) could be used to assess time variability of the radio remnant of \textit{Swift} short GRBs in our sample, for different circum-merger medium densities. \textit{Swift} GRBs used in this analysis are listed along the vertical axis. Circum-merger densities are indicated in the legend, in units of cm$^{-3}$.}
\label{fig:maxdistance}
\end{figure}

From Fig. \ref{fig:maxdistance} we see that the time variability of the radio remnants of recent GRBs can be detected to potentially very large distances due to the bright and fast rising radio flare. At later times, while the flare is still bright, its evolution slows down and is therefore the radio remnant's time variability is not detectable unless we have a reference observation from before the GRB.

\section{Conclusion}
\label{sec:conclusion}
We examined late-time radio observations to establish whether some of the short GRBs detected in the past by {\it Swift}-BAT with no afterglow identification and no distance measurement, could be BNS mergers similar to GW170817 located within 200\,Mpc. 

The detectability of the late-time radio flares associated with slow BNS ejecta depends on several source and environment properties. The most important parameter is the density of the circum-merger medium. Most GRBs in our sample, if observed in 2019 and had a pre-GRB reference image, would have detectable radio flares out to $\gtrsim500$\,Mpc for $n\gtrsim0.1$\,cm$^{-3}$, and out to $\gtrsim100$\,Mpc for $n\gtrsim10^{-2}$\,cm$^{-3}$ (see Fig. \ref{fig:maxdistanceref}). Establishing time variability of the radio emission to disentangle the radio counterpart from unrelated sources in the field is more challenging. For this case, we find that variability due to radio flares can be found with our fiducial Epochs I and II observations out to $\sim200$\,Mpc for $n\gtrsim10^{-1}$\,cm$^{-3}$ (see Fig. \ref{fig:maxdistance}). For comparison, measured circum-merger densities for cosmological short GRBs with measured afterglow fall in the range of $10^{-4}-1$\,cm$^{-3}$ \citep{2015ApJ...815..102F}. The estimated circum-merger density of GW170817 is around $5\times 10^{-3}$\,cm$^{-3}$ \citep{2018PhRvL.120x1103L}, at which density radio identification is challenging. In this case radio emission for older GRBs could be identified out to the distance of GW170817.

The above estimates depend also on the properties of the kilonova ejecta, in particular its outflow velocity. In our example of two velocity components, the slower, $0.1c$ has almost no effect on the radio emission in the first 20 years after the merger, even though it carries greater mass. It is possible that some component of the ejecta reaches velocities greater than our maximum $0.3c$, which would lead to faster rising and brighter radio emission. For example, \cite{2016ApJ...831..190H} considers an ejecta velocity distribution that extends to much higher velocities than $0.3c$. They find that radio emission peaks around $1-6$ years after the merger for $n=1-10^{-3}$\,cm$^{-3}$, compared to our $5-50$ years for the same $n$ range. For such higher velocities, more recent GRBs are more relevant for radio observations, and potentially lower $n$ values can also be probed. It will also be interesting to consider the detectability of long-term X-ray emission from the same interaction between the dynamical/wind ejecta and the circum-merger medium \citep{2018MNRAS.476.5621B}.

The discovery of radio remnants would confirm the BNS origin of a GRB, would provide valuable information on the BNS merger rate, the structure of the relativistic outflows, the properties of kilonova ejecta, and the density of the circum-merger medium. It could also trigger a deep GW search in historical data available for some of the GRBs. 

The authors are thankful for the generous support of the University of Florida, Texas Tech University and Columbia University in the City of New York. A.C. acknowledges support from NSF CAREER award 1455090. The Columbia Experimental Gravity group is grateful for the generous support of the National Science Foundation under grant PHY-1708028.

\section*{APPENDIX: GRB observations}
\label{sec:appendix}
Below we summarize the \textit{Swift} afterglow searches that have been carried out for the short GRBs we consider in this work. We also summarize the optical follow-up that has been carried out, and examine whether the obtained optical upper limits constrain kilonova emission from the source. To this end, we compare the optical constraints to the the observed kilonova light curve of GW170817 from \cite{2017ApJ...848L..17C} (their Fig. 1; see also \citealt{2017ApJ...848L..18N}), shifted from an assumed distance of 40\,Mpc to 200\,Mpc.  



\begin{itemize}
\item {\bf GRB\,170524A} was detected by \textit{Swift}/BAT \citep{GCN21135}. \textit{Swift}/XRT began observing 60.7\,s after the trigger, and found an uncatalogued X-ray source directionally coincident with the GRB \citep{GCN21135}. \textit{Swift}/UVOT began observations 67\,s after the trigger, but found no UV counterpart down to a limiting magnitude of 19.6 mag with its White filter \citep{GCN21135, GCN21141}. 
Several optical telescopes followed up the event, including the 0.5m robotic telescope D50 \citep{GCN21140}, Zeiss-1000 (East) of Tien Shan Astronomical Observatory and AZT-8 telescope of CrAO \citep{GCN21145}, and the Reionization and Transients Infrared Camera (RATIR) on the 1.5m Harold Johnson Telescope at the Observatorio Astronomico Nacional \citep{GCN21152}. These observations rule out a GW170817-like kilonova within 1\,Gpc. For example RATIR constrains the counterpart in the r band to $>25.66$\,mag 36 hours after the burst, at which time GW170817 at 200\,Mpc would have had an r band magnitude of 24.5. 
Radio follow-up was carried out by the AMI Large Array at 15 GHz beginning 0.2 days, 1.5 days, 3.4 days and 7.4 days after the burst that constrained the source flux density to 162 $\mu$Jy, 207 $\mu$Jy, 120 $\mu$Jy and 126 $\mu$Jy respectively \citep{GCN21212}. 
\item {\bf GRB\,170112A} was initially detected by \textit{Swift}/BAT \citep{GCN20436}. \textit{Swift}/XRT began observing 62.3\,s after the trigger, but found no X-ray afterglow \citep{GCN20436, GCN20440}. \textit{Swift}/UVOT began observing 66\,s after the trigger, but found no credible optical counterpart but found no UV counterpart down to a limiting magnitude of 19.6 mag with its White filter \citep{GCN20451}. The event was followed up by optical telescopes, including the 60-cm robotic telescope REM located at the La Silla Observatory \citep{GCN20438}, the MASTER robotic telescope \citep{GCN20439}, RATIR \citep{GCN20444}, and the AS-32 (0.7m) telescope of Abastumani Observatory \citep{GCN20472}. The most sensitive of these observations was RATIR's that reached 19 mag in the r, i, Z and Y bands about 3 hours after the burst. While we do not know the luminosity of the GW170817 kilonova at this early time, its peak luminosity expected at 200\,Mpc would be about 20.5 mag in these bands. These observations, therefore, cannot rule out a GW170817-like kilonova within 200\,Mpc. No radio follow-up was reported for this event. 
\item {\bf GRB\,160726A} was detected by \textit{Swift}/BAT \citep{GCN19731}, \textit{Fermi}/GBM \citep{GCN19732} and CALET Gamma-Ray Burst Monitor \citep{GCN19751}. \textit{Swift}/XRT and \textit{Swift}/UVOT follow-up observations were not carried out due to a Sun observing constraints \citep{GCN19731}. No optical or radio follow-up was reported for this event. 
\item {\bf GRB\,151228A} was detected by \textit{Swift}/BAT \citep{GCN18731} and Fermi/GBM \citep{GCN18736}.  \textit{Swift}/XRT and \textit{Swift}/UVOT follow-up observations were not carried out due to observing constraints \citep{GCN18731}. The 0.6-meter T60 telescope began observing 90\,s after the GRB, and detected no optical afterglow emission \citep{GCN18746}. The 0.5-meter MITSuME Akeno optical telescope searched for afterglow emission starting 16.3 hours after the trigger, but detected no emission \citep{GCN18747}. The latter, deeper observation reached a depth of 20 mag in the R and g bands, which cannot rule out a GW170817-like kilonova within 200\,Mpc. No radio follow-up was reported for this event. 
%
%
\item {\bf GRB\,141205A} was detected by \textit{Swift}/BAT \citep{GCN17137} and Fermi/GBM \citep{GCN17143}.  \textit{Swift}/XRT and \textit{Swift}/UVOT began observations 6.65 hours after the trigger, but found no afterglow emission \citep{GCN17141, GCN17144}. \textit{Swift}/UVOT reached a limiting magnitude of 20.7 in the u band. For comparison, the kilonova from GW170817 at 200\,Mpc would have had a u-band flux of 21.7 at 14.5 hours after the merger (based on \citealt{2017Sci...358.1565E}). We do not expect the kilonova light curve in the u band to be significantly brighter at 6.65 hours compared to 14.5 hours (e.g., Fig. 5 of \citealt{2017LRR....20....3M}), therefore we do not consider this as a constraint on kilonovae within 200\,Mpc. Optical follow-up observations have also been carried out by the 1\,m telescope located at Nanshan, Xinjiang, China, about 9 hours after the merger \citep{GCN17142}. No counterpart was found down to a limiting magnitude of 20.4 in the r band. This cannot rule out a GW170817-like kilonova within 200\,Mpc that would peak at 21 mag in the r band. No radio follow-up was reported for this event. 
\item {\bf GRB 140606A} was detected by \textit{Swift}/BAT \citep{GCN16353} \textit{Swift}/XRT began observing 1.3\,hrs after the trigger, but found no X-ray afterglow \citep{GCN16357}. \textit{Swift}/UVOT began observation 68\,s after the trigger, but found no optical afterglow down to a limiting magnitude of 20.8 in the u band \citep{GCN16358}. This observation was too early to detect a kilonova signal. Optical follow-up observations have been carried out by several observatories, including the 1m telescope located at Nanshan, Xinjiang, China \citep{GCN16359}, the 0.6m TELMA robotic telescope at the BOOTES-2 astronomical station Malaga, Spain \citep{GCN16361}, the AS-32 (0.7m) telescope of Abastumani Observatory \citep{GCN16371}, and the 6-meter BTA (+Scorpio-I) at SAO-RAS, Zelenchuk, Russia \citep{GCN16411}. The deepest observation by BTA 60 hours after the burst limited the r-band flux down to 22 mag. At 60 hours, a GW170817-like kilonova at 200\,Mpc would have an r-band flux of 22.3. Therefore we cannot fully rule out a GW170817-like kilonova within 200\,Mpc. No radio follow-up was reported for this event. 
\item {\bf GRB\,140414A} was detected by IPN \citep{GCN16110} and Swift/BAT \citep{GCN16110, GCN16111}. Swift/XRT began observing 11.2 hours after the trigger, but found no X-ray afterglow \citep{GCN16113}. Swift/UVOT began observation about 11 hours after the trigger, but found no optical afterglow down to a limiting magnitude of 20 with its white filter \citep{GCN16114}. This is below the expected flux of 21.7 mag in the narrower u-band of a GW170817-like kilonova at 200\,Mpc at 14.5 hours. Optical follow-up observations have been carried out by the DOLORES camera on the 3.6-m TNG Telescope at Canary Islands 15.2 hours after the burst, but found no optical counterpart down to a limiting magnitude of $\sim22$ in the r band \citep{GCN16112}. At 15.2 hours, a GW170817-like kilonova at 200\,Mpc would have an r-band flux of 20.8. Therefore, these observations rule out a GW170817-like kilonova within 200\,Mpc. 
%
%
\item {\bf GRB 130626A} was detected by Swift/BAT \citep{GCN14931}. Swift/XRT began observing 111 seconds after the trigger. It found a weak X-ray source, however, this source showed no sign of fading not fade, therefore Swift/XRT data shows no sign of an afterglow \citep{GCN14931, GCN14937, GCN15004}.  Swift/UVOT began observation 67 seconds after the trigger, but found no credible optical afterglow \citep{GCN14931, GCN14937}. Such early Swift/UVOT observations do not constrain kilonova emission. Optical follow-up observations have been carried out by RATIR 200 seconds after the GRB \citep{GCN14943}, and the 0.76-m Katzman Automatic Imaging Telescope (KAIT), located at Lick Observatory 97\,s after the GRB \citep{GCN14944}. No optical counterpart was found. These observations are too early for kilonova detection, therefore they cannot rule out a GW170817-like kilonova. No radio follow-up was reported for this event. 
\item {\bf GRB 120403A} was detected by Swift/BAT \citep{GCN13191}. Due to a Sun observing constraint, Swift/XRT, and Swift/UVOT could not search for afterglow emission from this event \citep{GCN13191}. The MASTER II robotic telescope began observing 9.8 hours after the trigger, but found no optical afterglow down to a limiting magnitude of 16 \citep{GCN13197}. The depth of this observation is not sufficient to rule out a GW170817-like kilonova at 200\,Mpc. No radio follow-up was reported for this event. 
\item {\bf GRB 120229A} was detected by Swift/BAT \citep{GCN12997}. Due to a Sun observing constraint, Swift/XRT, and Swift/UVOT could not search for afterglow emission from this event \citep{GCN12997}. The Magellan/Clay 6.5-meter telescope carried out observations at 9.6 and 33.6 hours after the burst, but found no optical counterpart \citep{GCN13084}. They report a limiting magnitude of 22.4 in the r-band at 9.6 hours after the burst. While no r-band observations of GW170817 were reported at this time, extrapolating later observations to this time from \cite{2017ApJ...848L..17C} indicate an expected r-band magnitude of 21.2 at 200\,Mpc. Therefore these observations are sufficient to rule out a GW170817-like kilonova at 200\,Mpc.
\item {\bf GRB 111126A} was detected by Swift/BAT \citep{GCN12599}. No follow up observation was carried out by Swift or others as the location of the event was about 29 degrees from the Sun \citep{GCN12599}.
%
%
%
\item {\bf GRB 101129A} was detected by IPN, Swift/BAT \citep{GCN11436}, Konus-Wind \citep{GCN11439}, and Suzaku WAM \citep{GCN11443}. Swift/XRT and Swift/UVOT began observing 11 hours after the trigger, but found no X-ray or optical afterglow emission \citep{GCN11436}. No limiting magnitude was reported for Swift/UVOT. No further optical follow-up was reported.
%
%
\item {\bf GRB 090417A} was detected by Swift/BAT \citep{GCN9133}. Due to an observing constraint, no Swift/XRT nor Swift/UVOT observations were carried out \citep{GCN9133}. A potential host galaxy was identified in the 2MASS catalog that was within the error region of the GRB, with its angular distance suggestive at roughly
97\%-confidence \citep{GCN9134}. The galaxy has a distance of about 400\,Mpc \citep{GCN9136}. No optical follow-up observation was reported. VLA carried out follow-up observations at 8.46\,GHz within a day of the GRB in the direction of the 2MASS galaxy, but found no counterpart down to a limiting flux of about 100\,$\mu$Jy \cite{GCN9160}.
%
%
\item {\bf GRB 080121} was detected by Swift/BAT \citep{GCN7209}. Swift/XRT and Swift/UVOT began observing 2.3 days after the trigger, but found no X-ray or optical afterglow emission \citep{GCN7224,GCN7217}.  Swift/UVOT reported a limiting magnitude of 20.73 in the v band. For comparison, at 2.3 days, a GW170817-like kilonova at 200\,Mpc would have a flux of 22.1 mag in the r-band, which should be brighter than the v band. Two nearby SDSS galaxies were identified within Swift/BAT error radius at a distance of $\approx 200$\,Mpc \citep{GCN7210}. Several other, fainter sources were also found to be present in the field, suggesting that the GRB may have occurred in a group or small cluster at this distance. No further optical or radio follow-up was reported.
\item {\bf GRB 070923} was detected by Swift/BAT \citep{GCN6818}. Swift/XRT and Swift/UVOT data is not available due to the GRB's direction having been too close to that of the Sun \citep{GCN6818}. VLA searched for but found no afterglow at a frequency of 8.46 GHz 5 days after the event, reaching a limiting flux density of 90\,$\mu$Jy \citep{GCN6831}.
\item {\bf GRB 070810B} was detected by Swift/BAT \citep{GCN6743}. Swift/XRT and Swift/UVOT began observing 60 seconds after the trigger, but found no X-ray or optical afterglow emission \citep{GCN6743, GCN6754, GCN6755}, although Swift/XRT reported two low-significance X-ray sources \citep{GCN6754}. A second observing epoch with Swift/XRT starting at 42 days detected an X-ray source within the Swift/BAT error circle \citep{GCN6852}. It was not reported whether this source faded with time. Swift/UVOT observations were too early to limit kilonova emission. Optical follow-up observations have been carried out by the KANATA 1.5-m telescope at Higashi-Hiroshima Observatory, Japan 3 minutes after the burst \citep{GCN6746}, the Xinglong TNT 80cm telescope 550\,s after the burst \citep{GCN6747}, the 2-m Faulkes Telescope South 3 minutes after the burst \citep{GCN6758}, the Shajn 2.6m telescope of CrAO \citep{GCN6762} and the Keck I telescope 13 hours and 6 days after the burst \citep{GCN6771}. The Shajn telescope observed a marginally detectable object, while others reported no observation. Using the deepest limit from Keck, assuming that kilonova emission at 6 hours is comparable to that at 12 hours after the burst, the obtained limit of 25.5 mag in the r band rules out a GW170817-like kilonova within 200\,Mpc. This galaxy is within the declination range within which observations are significantly degraded due to satellite transmission, therefore we do not include this GRB in Table \ref{table:GRBs}.
\item {\bf GRB 070406} was detected by Swift/BAT \citep{GCN6247}. Swift/XRT and Swift/UVOT began observing 20.6 hours after the trigger \citep{GCN6255, GCN6258}, but found no X-ray or optical counterpart \citep{GCN6255, GCN6262, GCN6263}. Swift/UVOT excluded a counterpart down to a limiting magnitude 22.4 in the u band. For comparison, at 20.6 hours, a GW170817-like kilonova at 200\,Mpc would have a flux of 21.5 mag in the u-band. Therefore, the Swift/UVOT observation rules out a GW170817-like kilonova within 200\,Mpc. 
%
%
\item {\bf GRB 051105A} was detected by Swift/BAT \citep{GCN4188}. Swift/XRT and Swift/UVOT began observing 67 seconds after the trigger, but found no X-ray or optical afterglow emission \citep{GCN4188, GCN4195, GCN4200}. The Swift/UVOT observation was too early to constrain kilonova emission. Optical follow-up has been carried out by the 14 inch Automated Response Telescope 2.8 hours after the burst \citep{GCN4189}, the Tautenburg 1.34-m Schmidt telescope 10 hours and 34 hours after the burst \citep{GCN4196, GCN4203}, TNG equipped with DOLORES 13 hours after the burst \citep{GCN4201}, the MDM 2.4m telescope 20 hours after the burst \citep{GCN4202}, the 1.0m Nainital telescope starting 8 hours after the burst \citep{GCN4204}, the 0.25m GETS telescope in the Gunma Astronomical Observatory 3 hours after the burst \citep{GCN4242}, and the 1.5m telescope of Maidanak Astronomical Observatory 7.6 hours after the burst \citep{GCN4349}. No plausible optical counterpart was reported. While for several of these observations no upper limit was reported, based on the magnitudes of the reported objects within the field of view, MDM seems to have had a depth down to $\sim23$ mag in the r band. At 20.6 hours post-burst when the MDM observation was carried out, a GW170817-like kilonova at 200\,Mpc should have an r-band magnitude of 20.8. Therefore, the MDM limits likely rule out a GW170817-like kilonova at 200\,Mpc. 
\item {\bf GRB 050925} was detected by Swift/BAT \citep{GCN4034}. Swift/XRT and Swift/UVOT began observing 91 and 92 seconds after the trigger, respectively, but found no X-ray or optical afterglow emission \citep{GCN4034, GCN4038, GCN4043}. The Swift/UVOT observation was too early to constrain kilonova emission. Optical follow-up has been carried out by the 2-m Faulkes North Telescope 3.3 minutes after the burst \citep{GCN4035}, the 0.8-m telescope at XingLong Observatory, China 2.2 hours after the burst \citep{GCN4036}, the PAIRITEL 1.3m telescope 18 hours after the burst \citep{GCN4042}. No optical counterpart was found. The first two of these follow-ups were too early to constrain kilonova emission. The PAIRITEL observation reached a limiting magnitude of $\approx18.5$ in the K and H bands, below the expected $\approx21.5$ mag in these bands for a GW170817-like kilonova emission from 200\,Mpc. 
%
\item {\bf GRB 050906} was detected by Swift/BAT \citep{GCN3926}. Swift/XRT began observing 79 seconds after the trigger, but found no X-ray or optical afterglow emission \citep{GCN3926, GCN3934}. Swift/UVOT was in safe mode and could not observe the source \citep{GCN3926}. Optical follow-ups have been carried out by the Robotic Palomar 60-Inch Telescope 114 seconds after the burst \citep{GCN3931}, the AEOS Burst Camera (ABC), attached to the AEOS telescope 15 minutes after the burst \citep{GCN3936}, FORS2 on the 8.2-m Antu Telescope at ESO/Paranal about a day after the burst \citep{GCN3940}, and the Russian-Turkish 1.5-m telescope 13 hours after the burst \citep{GCN3955}. No optical counterpart was found. VLT reached a limiting magnitude of 24.5 in the r band, compared to the 20.8 mag expected from a GW170817-like kilonova at 200\,Mpc. Therefore, VLT observations rule out a GW170817-like kilonova within 200\,Mpc. 
\item {\bf GRB 050202} was detected by Swift/BAT \citep{GCN3005}. No Swift/XRT or Swift/UVOT observations have been reported. The Swope 40-inch telescope at Las Campanas Observatory began observations 6 hours after the trigger, but found no optical afterglow emission down to a limiting magnitude of 16 in the r band \citep{GCN3006}. The 0.6\,m telescope at Mt. John University
Observatory began observations 12.6 hours after the trigger, but found no optical counterpart down to a limiting magnitude of about 20 in the r band \citep{GCN3018}. For comparison, the r band magnitude of a GW170817-like kilonova at 200\,Mpc 12 hours after the merger is 21. The VLA carrier out two epochs of radio observations but found no radio afterglow emission \citep{GCN3007, GCN3009}.
\end{itemize}

\bibliographystyle{mnras}
\bibliography{Refs}


\end{document}